\documentclass{vgtc}                          

\ifpdf
  \pdfoutput=1\relax                   
  \usepackage{graphicx}                
  \DeclareGraphicsExtensions{.pdf,.png,.jpg,.jpeg} 
\else
  \usepackage{graphicx}                
  \DeclareGraphicsExtensions{.eps}     
\fi%

\graphicspath{{figures/}{pictures/}{images/}{./}} 

\usepackage{microtype}                 
\PassOptionsToPackage{warn}{textcomp}  
\usepackage{textcomp}                  
\usepackage{mathptmx}                  
\usepackage{times}                     
\usepackage{cite}                      
\usepackage{tabu}                      
\usepackage{booktabs}                  

\usepackage{xcolor}
\usepackage{tabularx}

\usepackage{setspace}

\newcommand{\name}{\textit{TradAO}}

\newcommand{\taylor}[1]{\textcolor{black}{#1}}
\newcommand{\haotian}[1]{\textcolor{black}{#1}}

\onlineid{1062}

\vgtccategory{Research}

\vgtcinsertpkg

\title{{\name}: A Visual Analytics System for Trading Algorithm Optimization}

\author{
Ka Wing Tsang, Haotian Li \thanks{ {kwtsangae, haotian.li}@connect.ust.hk.}\\
\scriptsize HKUST
\and
Fuk Ming Lam \thanks{tonylam@algogene.com.}\\
\scriptsize ALGOGENE
\and
Yifan Mu, Yong Wang
\thanks{{ymuaa, ywangct}@connect.ust.hk; Y. Wang is the corresponding author.}\\
\scriptsize HKUST
\and
Huamin Qu
\thanks{huamin@cse.ust.hk.}\\
\scriptsize HKUST
}

\vspace{-1em}
\teaser{

 \includegraphics[width=\linewidth]{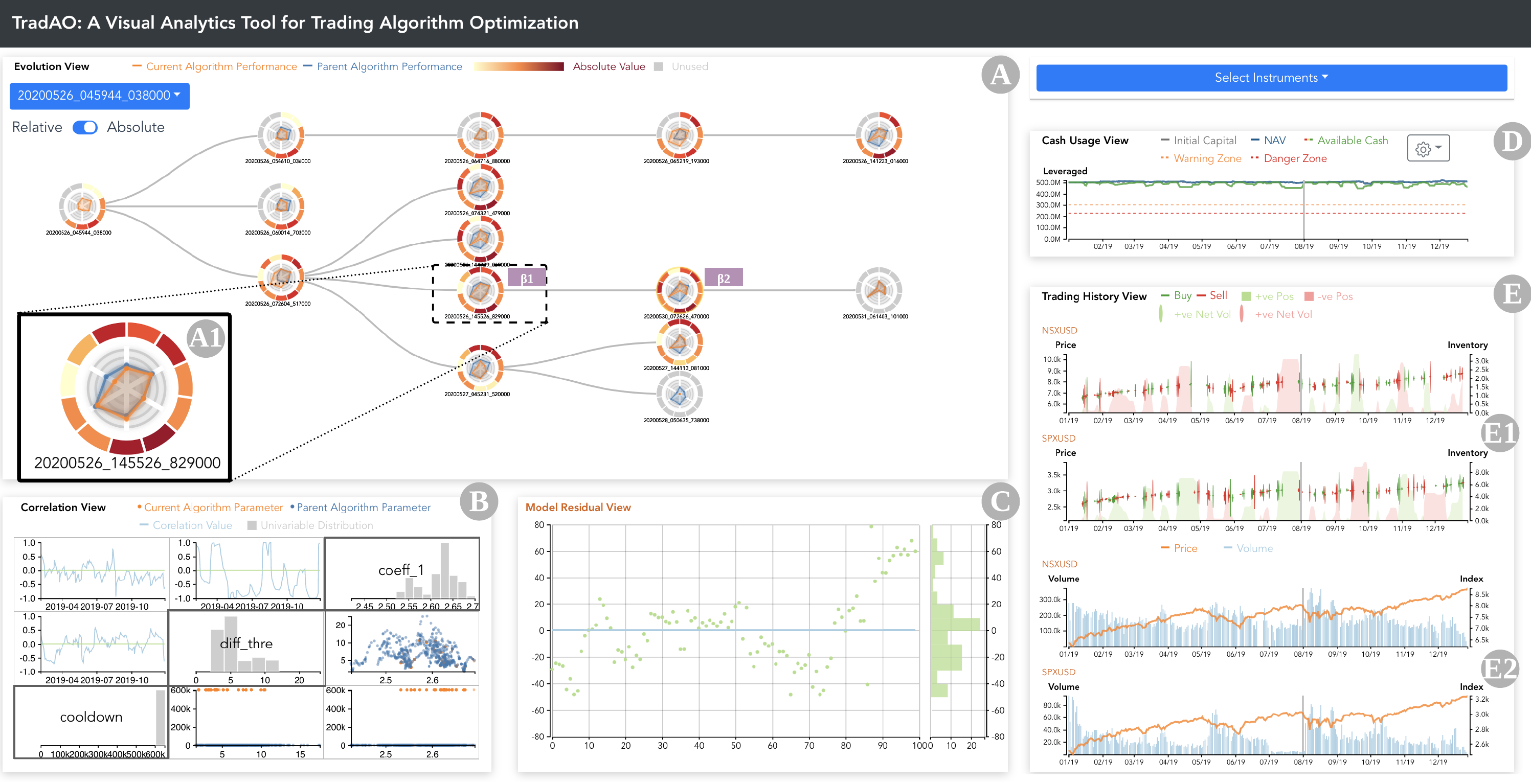}
 \caption{{\name} assists users in exploring the whole optimization process of a trading algorithm and evaluating its detailed performances. A) The algorithm evolution view explores how a trading algorithm evolves along with the configuration of different parameters. B) The parameter correlation view examines if a trading algorithm instance is overfitting. 
 \haotian{C) The trading residual view helps traders confirm the performance consistency of a trading algorithm instance more efficiently.} 
 D) The cash usage view facilitates monitoring the cash usage along different trading periods. E) The trading history view allows users to explore trading order information along the whole trading period.}
 \label{fig:teaser}
}

\abstract{
With the wide applications of algorithmic trading, it has become critical for traders to build a winning trading algorithm to beat the market. \haotian{However, due to the lack of efficient tools, traders mainly rely on their memory to manually compare the algorithm instances of a trading algorithm
and further
select the best trading algorithm instance for the real trading deployment.} We work closely with industry practitioners to discover and consolidate user requirements and develop an interactive visual analytics system for trading algorithm optimization. 
Structured expert interviews are conducted to evaluate {\name} and a representative case study is documented for illustrating the system effectiveness.
To the best of our knowledge, previous financial data visual analyses have mainly aimed to assist investment managers in investment portfolio analysis but have neglected the need of traders in developing trading algorithms for portfolio execution. {\name} is the first visual analytics system that assists users in comprehensively exploring the performances of a trading algorithm with different parameter settings.
} 

\CCScatlist{
  \CCScatTwelve{Human-centered computing}{Visu\-al\-iza\-tion}{Visualization design and evaluation methods;}{}
}

\begin{document}



\maketitle

\vspace{-0.5em}
\section{Introduction}
\haotian{Algorithmic trading, 
has been transforming the ﬁnancial market by utilizing computer algorithms in the trading over the recent two decades~\cite{intro_a}.
The time of doubling the total trading volume of stocks in Dow Jones Industrial Average has accelerated from 7.5 years to 2.9 years in the recent decades~\cite{intro_a}.}
Due to the increasing level of market fragmentation, both the trade execution cost and monitoring cost across markets are rising significantly~\cite{intro_c}. \haotian{Such a cost increase promotes algorithmic trading.}

\haotian{Building a winning trading algorithm to beat the market has become critical for traders to conduct successful trading.}
To evaluate the viability of a trading algorithm, traders need to conduct backtestings, where traders test the algorithm against historical data and observe its simulation performance.
Backtesting results are regarded as a predictive proxy for the potential future performance of an algorithm~\cite{intro_e}.
To build a winning trading algorithm, 
traders need to tune different algorithm parameters, conduct rounds of backtesting, and further compare and evaluate algorithm performances. 
Traders repeat this work ﬂow whenever a new set of trading instruction parameters and model variables is used. 
Such a process is often time-consuming and tedious~\cite{intro_f}.
Due to the lack of efficient tools, traders encounter multiple pain points in optimizing trading algorithms. 
First, it is challenging to compare \textit{algorithm instances} along the whole development process of a trading algorithm. An algorithm instance refers to a particular set-up of parameter combination. Traders can only do manual comparisons based on their memory. Second, the work efficiency of formulating inner parameters and variables of a trading algorithm is low and subjected to traders’ experience and intuition. Third, it is inconvenient for traders to trace 
the transaction details
of a trading algorithm in execution.                                                                                                  

To address these challenges, we propose {\name}, a novel visual analytics system to assist traders in  
exploring the trading algorithm optimization process
and evaluating its detailed performances. \haotian{Speciﬁcally, {\name} offers five well-coordinated views: \textit{Algorithm Evolution View} to overview a trading algorithm intuitively, \textit{Parameter Correlation View} to investigate the relationship between variables used in an algorithm instance, \textit{Trading Residual View} to examine the trading model performance consistency of an algorithm instance, \textit{Cash Usage View} to monitor the cashflow stability and\textit{ Trading History View} to evaluate the buying and selling pattern.} We designed the system with an experienced quantitative researcher who is also a co-author of this paper.
We conducted three structured expert interviews with domain experts to assess the usefulness and usability of {\name}. The major research contributions of this paper are summarized as follows:
\vspace{-0.5em}
\begin{itemize}
    \item A novel visual analytics system called {\name}, allowing traders to explore and compare different stages of a trading algorithm along the development cycle; 
    \vspace{-0.5em}
    \item \haotian{Preliminary} evaluations on the usefulness and usability of {\name} through structured interviews with domain experts, including both traders and system developers.
\end{itemize}

\section{Related Work}
The related work of this paper can be categorized into two groups: \textit{trading models} and \textit{financial data visualization}.

Extensive research has been conducted on algorithmic trading and various trading models have been proposed, for instance, \textit{Pairs Trading}~\cite{related_c, related_d}, \textit{Moving Average}~\cite{related_e, related_f}, \textit{Linear Regression}~\cite{related_g, related_h}, \textit{Neural Networks}~\cite{related_i, related_j} and \textit{Sentiment Analysis}~\cite{related_k, related_l}. Trading algorithms built on these trading models might involve complicated trading rules in an attempt to achieve a higher prediction accuracy. 
In this paper, we select three commonly used foundational trading models, which are relatively easier to comprehend yet quantitatively enough as being an entry point to evaluate our system, for demonstration. \textit{Pairs Trading} is a common market neutral strategy that opens long and short position for two securities simultaneously to capture the price differential~\cite{related_c}. \textit{Moving Average} usually applies to univariate time series for studying the relationship between a variable and its lagged terms~\cite{related_e}, while \textit{Multiple Linear Regression Model} is used for modelling multivariate time series~\cite{related_h}. 

Visual analytics methods have been  employed to assist financial industry practitioners in facilitating their
investment decision makings. For example,
FinanVis~\cite{related_m} is proposed to explore evolution patterns from financial news. WireVis~\cite{related_n} helps anti-money laundry specialist in bank to detect account suspicious transactions. FinVis~\cite{related_o} assists retail investors in doing personal finance planning. In addition to visualization systems assessing the overall situation in the market~\cite{related_p, related_q, related_r, related_s, related_t}, previous visualization systems also focused on portfolio analysis~\cite{related_u} for portfolio managers~\cite{related_t}. 
However, 
no visualization system exists for assisting traders in developing trading algorithms to the best of our knowledge.

\section{Task Analysis}
\label{sec-tasks}
We worked closely with four professional domain experts (D1, E1, E2, E3) from the financial field during the development of {\name}.
D1 is a quantitative researcher from ALGOGENE\footnote{\href{https://algogene.com/}{https://algogene.com/}} and also a co-author of this paper. ALGOGENE is an online trading algorithm development cloud platform.
E1 and E2 are a fund manager and a trader from a multi-asset management company and an investment bank respectively. E3 is a system interface developer in an investment bank. We conducted structured interviews with each expert and collected their feedback on the tasks and challenges of optimizing trading algorithms. \haotian{We further grouped the trading algorithm optimization tasks into three categories: \textit{strategy overview}, \textit{algorithm instance inspection} and \textit{algorithm instance assessment}.}  


\haotian{\textbf{Strategy overview}} 
tasks offer a full picture of the whole trading algorithm optimization cycle and enable inter-comparisons among algorithm instances with different parameter settings.

\textit{T1. Explore how a trading algorithm evolves over time.} 
\haotian{Developing an effective trading algorithm involves multiple rounds of backtesting with different parameter settings.
Traders are interested to gain a convenient overview of the model instances with different settings and their performances. }

\textit{T2. Identify how the effectiveness of a trading algorithm changes
under new parameter settings.}
The changes of model parameter settings
can significantly influence the effectiveness of different trading algorithms.
A common method to evaluate their effectiveness is to compare different performance metrics.

\haotian{\textbf{Algorithm instance inspection}}
tasks focus on the relationships among different parameters for a particular trading algorithm and enable inner-comparisons among those parameters. 

\textit{T3. Examine if a trading algorithm is overfitting.} A trading algorithm can be overfitting, which makes the algorithm achieve an ideal performance under a chosen time range in the backtesting simulation environment, but may perform badly on other time ranges. 
\haotian{Traders often need to check if a trading algorithm is overfitting by investigating the correlation between different parameters.}


\textit{T4. Verify if the trading algorithm
is robust.} A trading algorithm may be derived from economy theory or mathematical models. 
\haotian{To evaluate the appropriateness and robustness of an algorithm, traders rely on residual analysis to validate the underlying assumptions.}

\haotian{\textbf{Algorithm instance assessment}}
tasks reveal the viability of an algorithm
instance with specific parameter settings in terms of helping traders identify trading patterns in the real application. 
\haotian{Trading pattern refers to the transaction information such as the transaction time, frequency, price and volume, which can result in the fluctuation of the portfolio value over a trading period.}


\haotian{\textit{T5. Measure how well a trading algorithm optimizes the usage of investment capital.}
To maintain the fund as a self-financing portfolio for cash flow stability, it is crucial to strike a balance between leveraging positions and reserving cash buffer. Traders 
will record
the daily investment balance to monitor cash flows.}

\textit{T6. Assess the performances of a trading algorithm by comparing them with the market index.}
\haotian{Traders also evaluate a trading algorithm's effectiveness through its buying and selling patterns. A significant discrepancy between its trading pattern and the market trend can guide the subsequent refinement of a trading algorithm.}

\haotian{Regarding \textit{T2}, we select a set of performance metrics based on our discussion with domain experts. Specifically, there are six types of metrics that are often chosen by domain experts as a quick proxy to evaluate the overall performance of an algorithm instance, including Activeness, Consistency, Prediction, Profitability, Recovery and Robustness.
Activeness delineates how frequently an algorithm execute transactions; Consistency means the stability of an algorithm in making profits; Prediction represents the prediction accuracy of an algorithm; Profitability is the overall return of an algorithm; Recovery assesses how fast an algorithm recovers from previous losses; Robustness measures the stress tolerance level of an algorithm.
Specifically, we propose using nine measures belonging to the six types to evaluate the overall performance of a trading algorithm instance:
Mean Annualized Return (Yield)~\cite{visual_d} and Maximum Drawdown Percentage (MD)~\cite{visual_i} under the profitability category measure the profitability of an algorithm instance; Annualized Sharpe Ratio (Sharpe)~\cite{visual_f} and Annualized Sortino Ratio (Sortino)~\cite{visual_g} under the consistency category assess the risk adjusted return; Maximum Drawdown Duration (MaxDD)~\cite{visual_i} and Average Drawdown Duration (AvgDD~\cite{visual_i}) under the recovery category delineate the recovery time from loss. 99\% 1-day Return Value-at-Risk (VaR99)~\cite{visual_h} and Annualized Return Volatility (Vol)~\cite{visual_j} under the robustness category indicate the performance volatility; and Win Rate (WinRate)~\cite{visual_e} under the prediction category represents the prediction accuracy.}

\section{Visual Design}

\haotian{{\name} is a web-based application consisting of five main views: algorithm evolution view (Figure \ref{fig:teaser}(A)), parameter correlation view (Figure \ref{fig:teaser}(B)), trading residual view (Figure \ref{fig:teaser}(C)), cash usage view (Figure \ref{fig:teaser}(D)), and trading history view (Figure \ref{fig:teaser}(E)).}
The algorithm evolution view provides a quick overview of the performances and parameter settings of different algorithm instances. The parameter correlation view and the trading residual view then return the detailed parameter information of any selected algorithm instance on demand. Meanwhile the cash usage view and the trading history view display the trading patterns of the selected algorithm instance. 
In this paper, we analyzed the backtesting records collected by ALGOGENE platform. \haotian{These backtesting records were imported into {\name} through REST APIs.
The trading period of an algorithm 
is often several months or even several years, 
 while its trading frequency can be daily, hourly or even higher frequency~\cite{visual_a}. 
There is no fixed rule in deciding
the backtesting trading period, as it depends on traders' own preferences and the availability of backtesting platform's historical data.}
{\name} enables linked analysis across different views to facilitate convenient exploration of such records and help traders quickly find a good trading algorithm with appropriate parameter settings. The detailed visualization designs will be introduced in this section.

\subsection{Algorithm Evolution View}

\haotian{We provide the algorithm evolution view for users to explore how a trading algorithm evolves along with the configuration of different parameters before deployment (\textbf{T1}).}


The evolution of 
a trading algorithm is visualized as a tree diagram that consists of linked sequential glyphs. \haotian{Each glyph represents a trading algorithm instance, i.e., a trading algorithm with a specific parameter setting configured by a trader.
As shown in Figure~\ref{fig:teaser}(A1), each glyph consists of a outer ring and two inner radar charts.
Each segment of the outer ring represents 
a trading algorithm parameter.}
The color of the segment encodes the relative value of an algorithm parameter compared with the value of the same parameter used in other trading algorithm instances over the whole algorithm development.
A darker color corresponds to a higher parameter value.
\haotian{The inner radar chart visualizes the performance of an algorithm instance with respect to the six measure categories mentioned in section~\ref{sec-tasks}.}
Customized methods are used to normalize and standardize the characteristic scores into a scale from 0 to 100 for each dimension.
\haotian{The orange star plot shows the performance of the current algorithm instance, while the blue star plot indicates the performance of its parent algorithm instance. Parent algorithm instance refers to the algorithm instance that the current algorithm instance is derived from by further tuning its parameters.}
Users can make a quick comparison on the performance of two consecutive instances at a glance (\textbf{T2}). \haotian{The root glyph represents
the trading algorithm instance with initial parameter configurations
and the subsequent glyphs represent other algorithm instances that are further developed. }

\begin{figure}
    \centering
    \includegraphics[width=\linewidth]{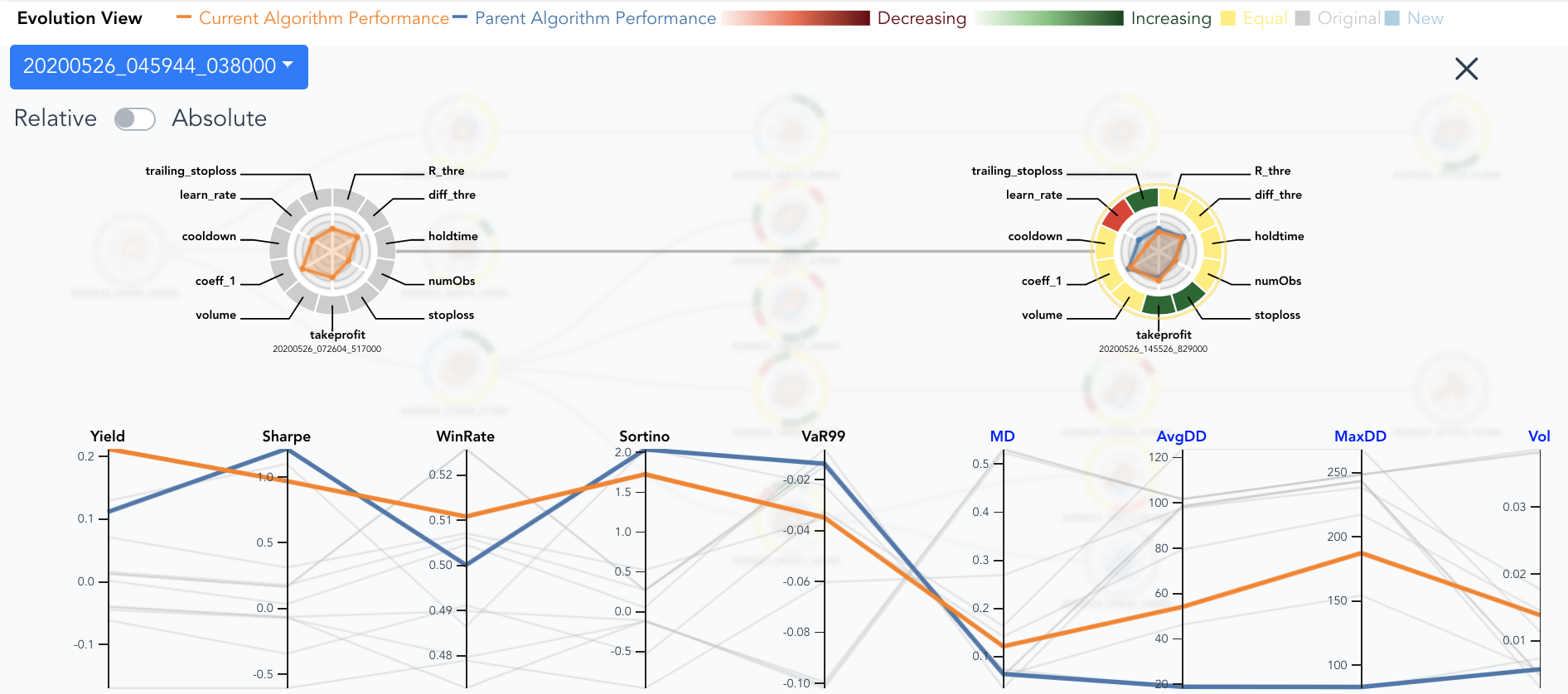}
    \caption{Performance comparison of two trading algorithms. \taylor{The higher the left five performance metric values (Yield, Sharpe, WinRate, Sortino, VaR99) on the parallel coordinate are, the better the trading algorithm performs. The lower the right four performance metric values (MD, AvgDD, MaxDD, Vol) are, the better the algorithm performs.}}
    \vspace{-2em}
    \label{fig:overlay}
\end{figure}

The algorithm evolution view enables rich interactions. 
First, when a user \taylor{hovers} over a segment on the outer ring or any spoke on the radar chart, a tooltip will display both the corresponding parameter name and value. 
Second, \taylor{a relative mode} offers users another perspective to overview the whole trading strategy and compare two consecutive trading algorithms through a different color scheme denoting the change of value of variables. The greener the segment is, the higher amount of the variable value increased in a particular trading algorithm compared to its parent. While the redder the segment is, the higher amount of the value decreased. 
\taylor{Third, when users click a particular glyph, 
an expanded view, consisting of a parallel coordinate and glyphs showing the current algorithm instance and its parent, 
will be overlaid in Evolution View, as shown in Figure~\ref{fig:overlay}.}
Users can evaluate the effectiveness of the selected trading algorithm instance and compare with other algorithm instances’ performance via the parallel coordinate. 
\taylor{Similar to the color encoding of the star plot,
the orange line in the parallel coordinate represents the selected algorithm instance while the blue line represents its parent. Grey lines representing all other algorithm instances are shown for benchmarking purposes.}
\taylor{When users click on a particular glyph, further details of that trading algorithm instance will be be shown in all the other views simultaneously.} 

\vspace{-0.5em}
\subsection{Parameter Correlation View}


The parameter correlation view helps users examine if a trading algorithm is overfitting (\textbf{T3}).
As shown in Figure~\ref{fig:teaser}(B),
it is a $3\times3$ grid, which can be categorized into three parts: lower right cells, diagonal cells and upper left cells. 
\haotian{The lower right cells shows three scatter plots 
that visualize the correlation between each pair of parameters used in both the selected algorithm instance (the orange scatter plot) and its parent algorithm instance (the blue scatter plot).  
The diagonal cells show univariable distribution histograms. 
The horizontal and vertical axes encode the value and occurrence probability of a variable used in the selected algorithm.}
The upper left cells show line charts to indicate the trends of correlation values over time. All the line charts’ vertical scale are normalized to
$[-1,1]$.
The green line in the middle denotes 0.
\haotian{Such a visualization provide users with deep insights into the correlation between parameters. The scatterplots in the bottom right offer traders a quick overview of the correlation, while the line charts enable a detailed observation of the correlation magnitude change over time.}


\subsection{Trading Residual View}
\haotian{The trading residual view helps traders confirm the performance consistency of a particular trading algorithm more efficiently (\textbf{T4}).
As shown in Figure~\ref{fig:teaser}(C), the trading residual view consists of a scatter plot in the left and a histogram in the right.
The scatterplot, where the horizontal axis represents the ordinal sequence of model residuals and vertical axis represents the residual value, shows the statistical properties of residuals, helping traders determine the confidence level of  generated trading signals at different time points.
The vertical histogram displays the probability distribution of residuals, where its horizontal axis is the discrete probability density and vertical axis is the domains of residual value. }


\vspace{-0.5em}
\subsection{Cash Usage View}
\haotian{The cash usage view (Figure~\ref{fig:teaser}(D))} \haotian{facilitates} monitoring the cash usage along different trading periods.
\taylor{It is a common practice that traders use leverage to increase the potential investment return. Leverage~\cite{leverage} is an investment approach through using borrowed capital.}
Liquidity~\cite{visual_k} is an important indicator to evaluate the viability of a trading algorithm (\textbf{T5}).
\taylor{Traders can use the cash usage view} to track both the net asset value (NAV) and remaining cash amount (Available Cash) along the whole trading period, which are represented by a blue line and a green line with dynamic colors, respectively.
The dotted grey line represents the initial amount of capital for users’ reference. Two benchmark lines warn users against the liquidity shortage. The orange dotted line represents the \textit{warning level} to indicate a potential liquidity risk, while the red dotted line 
indicates a significant liquidity risk (\textit{danger level}). 
When the remaining cash amount falls below the warning level or the danger level, the corresponding line becomes orange or red to alert users. 
Users can click on the top right gear button to interactively adjust the benchmark levels according to their own risk tolerance levels with the two sliders. 
Users can also use the brush and zoom function to drill into more cash usage details at a specific period and the trading history view will be updated simultaneously, helping to identify the discrepancy among the net asset value, cash flow and inventory flow.
\nopagebreak[0]

\vspace{-1em}
\subsection{Trading History View}
\haotian{The trading history view (Figure~\ref{fig:teaser}(E))} allows users to explore trading order information (e.g., the price and volume of a transaction) along the whole trading period. 
Users can use the trading history view to identify outlier transactions (\textbf{T5}). 
\haotian{In Figure~\ref{fig:teaser}(E1), horizontal axis represents the trading period of a trading algorithm. The left vertical axis represents the price of financial instruments and the right vertical  axis represents the inventory level involved in transactions.
The height of the ellipse represents the net volume transacted in a single trading day, while the center position of the ellipse is the average price of all transactions in a single trading day. 
Two green horizontal lines indicate the highest and lowest buying price when there are buying transactions. Two red horizontal lines indicate the highest and lowest selling price. }
The height of the shading in the background encodes the daily outstanding inventory amount. Green shading indicates a positive amount of inventory. Red shading refers to a negative amount of inventory, meaning a short position. 

\haotian{Users can also use this view to compare its trading patterns with the market trends of various financial instruments or market indices in Figure~\ref{fig:teaser}(E2) (\textbf{T6}). The Drop-down button on the top of the right panel enables users to interactively select displayed market charts.}



\section{Evaluation}
We conducted structured interviews with three domain experts to evaluate the usefulness and usability of {\name}.
Three domain experts (i.e., E1, E2 and E3), who participated in our task analysis (Section~\ref{sec-tasks}), are further interviewed to collect their feedback.
Each interview lasted for about 45 minutes and contained three sections, an introduction session to explain the project scope, the dataset and the visual encoding (10 mins), a task session to observe how interviewees use the system (20 mins), and a comment session to collect feedbacks in terms of system usability, effectiveness and visualization designs (15 mins). 


\subsection{Overall Feedback on {\name}}\label{overall_evaluation}
The overall system was confirmed to be useful for facilitating experts’ daily research and analysis tasks.
All experts appreciated the functionality of the algorithm evolution view in particular. E1 said that \textit{“It can help reduce a significant amount of workload in tracing my (trading algorithm development) changes, analyzing algorithm details and comparing different (algorithm instance) performance at the same time”}.
E2 confirmed that he would have overlooked some details and built a less preferable trading algorithm without the help of {\name}. In the task session, interviewees were asked to perform three practical tasks within a limited time period to examine the system effectiveness: \haotian{(1) select the “best” trading algorithm instance out of the given algorithm pool; (2) identify problematic algorithm configurations which may lead to a significant performance gap between two consecutive instances; and 
(3) suggest potential issues for the algorithm instances.
All the experts can finish all tasks within the given time and they confirmed the effectiveness of {\name}. }

\taylor{All the three experts said that the visualization designs are intuitive and easy to understand.} They appreciated the cross-view interactions, for example, enabling inline brushing in the cash usage view and the trading history view helps users quickly narrow down into the time periods of interest.

\vspace{-0.5em}
\subsection{Case Study: Identify the Problematic Algorithm Configurations}
We report a case found by E2 during our expert interviews to further demonstrate the usefulness of {\name}.
\haotian{To finish the second task, E2 first explored the evolution of a trading strategy based on \textit{Pairs Trading Model}
as shown in Figure~\ref{fig:teaser}(A) (\textbf{T1}). }
\haotian{E2 said that the tree-based visualization in the algorithm evolution view greatly
reduces his workload in narrowing down the scope of the instance candidates}. 

E2 noticed that $\beta1$ has the largest hexagon-like shape radar chart among all trading algorithm instances in Figure~\ref{fig:teaser}(A) and \haotian{confirmed that it was one of the optimal algorithms in terms of the highest Yield and Win Rate, as shown in Figure~\ref{fig:overlay}.
E2 found an upward NAV curve and green Available Cash curve in the cash usage view which indicated high profitability and cash flow stability (\textbf{T5}).} 
E2 speculated that $\beta1$ may be a good algorithm instance for deployment after a quick review on model parameters shown by its outer ring.

E2 was also interested to examine the descendent algorithm instances of $\beta1$. He made a quick comparison through the radar charts in $\beta2$ and noticed a significant performance gap between $\beta1$ and $\beta2$ (\textbf{T2}). To further explore the reasons behind this gap, he continued to explore the trading history view and spotted a significant change from a sizeable pink shading area to a green shading on NSXUSD (a financial instrument) in August 2019 as shown in Figure~\ref{fig:teaser}(E1). This reflected a substantial momentum change in holding inventory level from short position to long position. SPXUSD was traded reciprocally during the same period. E2 further checked Figure~\ref{fig:teaser}(E2) to confirm if these trade actions match the corresponding market direction (\textbf{T6}). \haotian{The percentage decrease in NSXUSD market index is greater than the drop in SPXUSD market index. E2 regarded $\beta2$ as an ineffective instance and investigated into its parameter details in the parameter correlation view as shown in Figure~\ref{fig:teaser}(B) (\textbf{T3}).
E2 observed the line charts and identified
a highly fluctuated correlation value pattern between Coeff\_1 
and Diff\_thre.} 
E2 cross-checked the trading algorithm robustness of $\beta2$ with the trading residual view as shown in Figure~\ref{fig:teaser}(C) (\textbf{T4}) and observed a non-random cyclic pattern. E2 speculated that $\beta2$ is statistically non-stationary and probably ineffective to generate good performance. 
\section{Conclusion}

We propose a novel visual analytics system, {\name}, for helping traders explore and compare different trading algorithm instances to find an appropriate trading algorithm. Expert interviews with professional traders and trading algorithm developers provide support for the usefulness and usability of {\name}.
However, our approach is not without limitations. 
\haotian{First, {\name} suffers from scalability issues.}
For instance,
when traders built and backtest the same trading algorithm for too many times,
it will be difficult to display all the algorithm
instances in the algorithm evolution view due to the space limit.
\haotian{Second, {\name} only provides a finite set of pre-selected performance metrics and may not satisfy different requirements of users under all conditions. 
Third, 
our current qualitative evaluation approach could be supplemented with an additional quantitative analysis to measure the system effectiveness in terms of time spent for task completion under a control set-up.
In future work, we would like to further improve {\name} by handling the above limitations.}


\bibliographystyle{abbrv-doi}

\bibliography{template}
\end{document}